\documentclass[twocolumn,showpacs,floats,floatfix,superscriptaddress,aps,prl]{revtex4}
\usepackage{amsfonts}
\usepackage{amssymb}
\usepackage{amsmath}
\usepackage{calc}
\usepackage{graphicx}
\usepackage{bm}

\def\be{ \begin{equation} }
\def\ee{ \end{equation} }
\def\bea{ \begin{eqnarray} }
\def\eea{ \end{eqnarray} }
\def\ba{ \begin{array} }
\def\ea{ \end{array} }
\def\bse{\begin{subequations}}
\def\ese{\end{subequations}}

\def\Osvet{\mathcal{O}}

\def\ket#1{\vert #1 \rangle}
\def\bra#1{\langle #1 \vert}
\def\ibid{\textit{ibid.\ }}
\def\sech{\text{sech\,}}
\def\etal{\textit{et al.}}
\def\vol{\textbf}
\def\PRA{Phys. Rev. A\ }

\def\W{\psi_W}
\def\i{\text{i}}
\def\e{\text{e}}

\def\ee{ \end{equation} }

\def\phonons{\mu}

\begin{document}

\author{Svetoslav S. Ivanov}
\affiliation{Department of Physics, Sofia University, James Bourchier 5 blvd, 1164 Sofia, Bulgaria}
\author{Peter A. Ivanov}
\affiliation{Department of Physics, Sofia University, James Bourchier 5 blvd, 1164 Sofia, Bulgaria}
\author{Nikolay V. Vitanov}
\affiliation{Department of Physics, Sofia University, James Bourchier 5 blvd, 1164 Sofia, Bulgaria}
\affiliation{Institute of Solid State Physics, Bulgarian Academy of Sciences, Tsarigradsko chauss\'{e}e 72, 1784 Sofia, Bulgaria}
\title{Quantum search with trapped ions}
\date{\today }

\begin{abstract}
We propose an ion trap implementation of Grover's quantum search algorithm for an unstructured database of arbitrary length $N$.
The experimental implementation is appealingly simple because the linear ion trap allows for a straightforward construction,
 in a single interaction step and without a multitude of Hadamard transforms, of the reflection operator, which is the engine of the Grover algorithm.
Consequently, a dramatic reduction in the number of the required \textit{physical} steps takes place,
 to just $\Osvet(\sqrt{N})$, the same as the number of the \textit{mathematical} steps.
The proposed setup allows for demonstration of both the original (probabilistic) Grover search and its deterministic variation,
 and is remarkably robust to imperfections in the register initialization.
\end{abstract}

\pacs{03.67.Lx, 03.67.Ac, 03.67.Bg, 42.50.Dv}
\maketitle

{\bf \em Introduction.}
Quantum computers hold the promise of a dramatic speed-up for certain types of problems, which are processed intrinsically slowly by classical computers \cite{Nielsen}.
The fast search invented by Grover \cite{Grover} is arguably the quantum algorithm with the most outreaching implications.
The Grover algorithm searches an arbitrary element in an unsorted database with $N$ entries quadratically faster than its classical counterpart, with only $\Osvet(\sqrt{N})$ calls to an oracle.
It is realized with the repeated application of two operations on an initial uniform superposition of all states
 -- an oracle, which flips the phase of the marked element $\ket{\psi_m}$, and a reflection of the state vector about the mean.
As $N$ increases, the fidelity approaches unity, with error $\Osvet(1/N)$.

Grover's algorithm has been demonstrated experimentally with two $(N=4)$ \cite{NMR-Grover} and three ($N=8$) \cite{NMR-8} qubits in nuclear magnetic resonance,
 and with $N=32$ items in classical optics \cite{Bhattacharya}.
This latter experiment demonstrated that entanglement is not essential for Grover's search but only the superpositional nature of the data register \cite{Lloyd}.

Trapped ions present one of the most plausible physical implementations of a quantum computer for their quantum dynamics is readily controlled and characterized and the system is scalable in principle.
The theory \cite{Cirac-Zoller} and a number of landmark trapped-ion experiments \cite{ion traps,W} have been a major driving force behind quantum computing in recent years.
Grover's search, however, has been implemented with a pair of ion qubits ($N=4$) only \cite{Brickman},
 largely because of the complexity of an ion-trap search experiment stemming from the necessity to construct the reflection operator,
 the standard implementation of which requires numerous quantum (mainly Hadamard) gates.

The reflection operation used in Grover's algorithm is known in matrix theory as \textit{Householder reflection} (HR) and is widely used in classical data analysis.
We have recently advocated the use of HR as a very efficient tool for a range of quantum-state engineering problems
 in single-particle \cite{Ivanov06,Ivanov07} and many-particle systems \cite{Ivanov08}.
These include the synthesis of arbitrary unitary transformations \cite{Ivanov06}, superposition-to-superposition transitions \cite{Ivanov07},
 and engineering of arbitrary mixed states \cite{Ivanov07} of a single particle,
 and quantum-state engineering of collective entangled states of trapped ions \cite{Ivanov08}.
This work has been triggered by the fact that the HR operator emerges naturally as the propagator in a coherently-driven degenerate two-level system:
 as a single HR for an $N$-pod system of one degenerate and one non-degenerate level \cite{Kyoseva06},
 and as a product of HRs in a system of two degenerate levels \cite{Kyoseva08}.

In this Letter, we make use of the HR implementation with collective states of trapped ions \cite{Ivanov08}
 to propose a very simple experimental demonstration of Grover's search. 
The linear ion-trap architecture and the ensuing linkage pattern prove ideally suitable for Grover's search,
 because the required HR operator arises naturally, as a solution of the Schr\"{o}dinger equation for a Hamiltonian inherent for laser-driven trapped ions.
Hence the global HR operator, used to perform the reflection about the mean,  can be produced in a \textit{single} interaction step, rather than by $\Osvet(N)$ gates hitherto,
 which greatly reduces the required overall nonquery resources \cite{Grover02}.

As stressed by Grover \cite{Grover02}, the total number (query and nonquery) of physical steps in implementations of the reflection operator
 with a multitude of Hadamard gates can exceed considerably the number of logical steps.
For example, in one of the most sophisticated experiments so far \cite{NMR-8}, about 100 pulses have been used to simulate Grover's search in an $N=8$ register.
The implementation proposed here is the first one, where the total number of \textit{physical} steps is the same as the number of algorithmic steps,
 which is $N_G =\left[\pi/(2\sin^{-1}(2\sqrt{N-1}/N))\right]$ ($[n]$ denotes the integer part of $n$), or $N_G =[(\pi/4)\sqrt{N}]$ for large $N$.
This speed-up reduces significantly the experiment complexity, the overall processing time, and the fidelity requirement for each interaction step.

\begin{figure}[tb]
\includegraphics[angle=0,width=50mm]{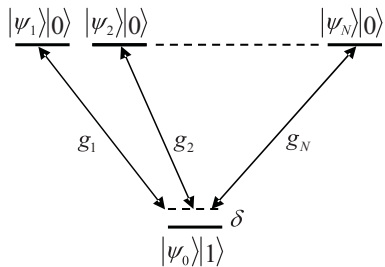}
\caption{Linkage pattern of the collective states of $N$ trapped ions driven by a red-sideband laser pulse.
The lower state $\ket{\psi_0}\ket{1}$ involves one vibrational phonon and no ionic excitations,
 whereas the upper set of states $\{\ket{\psi_k}\ket{0}\}_{k=1}^N$ involves zero phonons and a single ionic excitation of the $k$th ion.
}
\label{fig1}
\end{figure}

{\bf \em Hamiltonian.}
We consider a string of $N$ two-state ions confined in a linear Paul trap.
The radial trap frequencies are much larger than the axial frequency $\omega_z$, so that the ions form a linear string along the $z$ axis \cite{James}.
The ions are laser-cooled into their motional zero-phonon ground state $\ket{0}$ \cite{King} and
 their internal dynamics is controlled by a laser (or two lasers for a Raman transition), with a carrier frequency tuned near the red-sideband resonance:
 $\omega_{ n } = \omega_0 -\omega_z + \delta_{ n }$, where $\omega_0$ is the Bohr transition frequency of the ions
 and $\delta_{ n }$ is the detuning of the $n$th ion ($\vert \delta_n \vert \ll \omega_z$).
In the Lamb-Dicke limit and the rotating-wave approximation, the interaction Hamiltonian is \cite{James}
\be \label{Hamiltonian}
\mathbf{H}_I (t) = \hbar \sum_{n=1}^N \frac{\eta_n \Omega_n (t)}{2\sqrt{N}}\left[ a^{\dag}\sigma_n^- \e^{\i(\delta_nt + \phi_n) } + \text{h.c.} \right].
\ee
Here $a^{\dag }$ and $a$ are the creation and annihilation operators of the center-of-mass phonons, while
 $\sigma_n^+ = \ket{1_n} \bra{0_n}$ and $\sigma_n^- = \ket{0_n} \bra{1_n} $ are the Pauli spin-flip operators
 for the internal states $\ket{0_n}$ and $\ket{1_n}$ of the $n$th ion.
 $\eta _{n}=\sqrt{\hbar k_{n}^{2}\cos ^{2}\theta _{n}/2M\omega _{z}}$ is the Lamb-Dicke parameter,
 where $k_{n}$ is the laser wavevector, $\theta_{n}$ is the angle between the trap axis and the laser propagation direction,
 $M$ is the ion mass, 
 and $\Omega _{n}\left( t\right)$ is the time-dependent real-valued Rabi frequency of the laser-ion coupling.
All Rabi frequencies must have the same time-dependent envelope $f\left( t\right)$, but they may have different amplitudes $\Omega_{n}$.

Since the Hamiltonian \eqref{Hamiltonian} is of Jaynes-Cummings type, it conserves the sum of ionic and vibronic excitations;
 hence the Hilbert space decomposes into subspaces with a definite number of excitations.
The \textit{single-excitation subspace} is spanned by the $N+1$ basis states,
\begin{subequations}\label{Hilbert space}
\begin{eqnarray}
\ket{\psi_k}\ket{0} &=& \ket{ 0_1 \ldots 0_{k-1},1_k,0_{k+1}\ldots 0_N} \ket{0}, \label{k} \\
\ket{\psi_0}\ket{1} &=& \ket{0_1\ldots 0_N} \ket{1},  \label{0}
\end{eqnarray}
\end{subequations}
where  $\ket{\phonons}$  is the vibrational state with  $\phonons$  phonons  $(\phonons=0,1)$,
 and $\ket{ 0_1 \ldots 0_{k-1},1_k,0_{k+1}\ldots 0_N}$ is a collective ionic state ($k=1,2,\ldots N$), in which the $k$th ion is in state $\ket{1}$ and all other ions are in state $\ket{0}$.
Our quantum memory register will be the set of states \eqref{k}: $\left\{\ket{\psi_k}\right\}_{k=1}^N$;
 throughout the algorithm operation the population remains in vibronic state $\ket{0}$; it will be omitted hereafter.

{\bf \em Householder reflection.}
After a phase transformation the Hamiltonian \eqref{Hamiltonian} can be written as \cite{Ivanov08}
\begin{equation}
\mathbf{H}_I (t) = \frac{\hbar}{2} \sum_{n=1}^N  g_n (t) \ket{\psi_0}\bra{\psi_n} + \frac{\hbar}{2}\delta \ket{\psi_0}\bra{\psi_0} + \text{h.c.},
\label{H1}
\end{equation}
where all detunings are assumed equal, $\delta_n=\delta $,
 and $g_n(t) =\eta_n\Omega_n e^{-i\phi_n} f(t)/\sqrt{N} = g_n f(t)$ is the coupling between the internal and motional degrees of freedom for each ion ($n=1,2,\dots,N$).
The linkage pattern for this Hamiltonian is an $N$-pod, wherein state $\ket{\psi_0}$ is coupled to each state in the manifold $\left\{\ket{\psi_k}\right\}_{k=1}^N$, as shown in Fig.~ \ref{fig1}.
It has been shown very recently that the propagator within the single-excitation manifold $\left\{ \ket{\psi_k} \right\} _{k=1}^N$ for exact resonance ($\delta=0$)
 and for root-mean-square (rms) pulse area $A=g\int_{-\infty }^{\infty }f(t) dt = 2(2l+1) \pi$ (with $l=0,1,2,\ldots $), where $g^2=\sum_{n=1}^{N}\vert g_{n}\vert ^2$,
 is given by a \textit{standard HR} \cite{Ivanov06,Ivanov07,Ivanov08},
\begin{equation}
\mathbf{M}(\chi) =\mathbf{1}-2\ket{\chi} \bra{\chi},  \label{SHR}
\end{equation}
where the components of the $N$-dimensional normalized vector $\ket{\chi}$ are the normalized couplings,
\begin{equation}
\ket{\chi} =\frac{1}{g}\left[ g_1,g_2,\ldots,g_N\right] ^T.  \label{vectors}
\end{equation}

The \textit{generalized HR} within the manifold $\left\{ \ket{\psi_k} \right\} _{k=1}^N$,
\be
\mathbf{M}(\chi ;\varphi) = \mathbf{1} + (\e^{\i\varphi}-1) \ket{\chi} \bra{\chi},\label{GHR}
\ee
is realized at suitably chosen detunings.
For example, for a hyperbolic-secant pulse, $f(t)=\sech (t/T)$, with rms area $A = 2\pi l$ ($l=1,2,\ldots $),
 the HR phase $\varphi$ is produced by detunings, obeying the equation $2\arg \overset{l-1}{\underset{j=1}{\prod }} [\delta T+\i(2j+1)]=\varphi$ \cite{Kyoseva06}.
The generalized HR can be created also for other pulse shapes, e.g. Gaussian, but the required pulse area and detuning have to be evaluated numerically.

\begin{figure}[tb]
\centering  \includegraphics[angle=0,width=75mm]{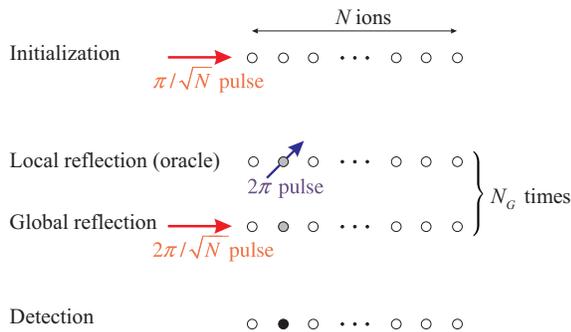}
\caption{(Color online) Implementation of Grover's search with $N$ trapped ions.
The ion chain is initialized in an entangled $W$-state by preparing it initially in state $\ket{\psi_0}\ket{1}$, and then apply a red-sideband pulse
 with a temporal area of $\pi/\sqrt{N}$ uniformly at all ions.
Then we perform repeatedly, $N_G$ times, a local reflection of the marked ion by applying a $2\pi$ pulse on it,
 followed by a global reflection, inflicted by a $2\pi/\sqrt{N}$ pulse applied to the entire ion string.
Finally, the marked ion is identified for it is the only one in state $\ket{1}$.}
\label{fig2}
\end{figure}

{\bf \em Grover search.}
The implementation of the Grover algorithm is depicted in Fig.~\ref{fig2}.

1. {\em Initialization}
We first prepare the register \eqref{k} in an equally weighted superposition of all $N$ states \cite{Grover},
 i.e., $\ket{\W} = \sum_{n=1}^N \ket{\psi_n}/\sqrt{N} = [ 1,1,\ldots,1 ]^T / \sqrt{N}$;
 this is the well-known entangled $W$-state, which has been demonstrated experimentally with eight ions \cite{W}.
For this, one can choose between the latter technique \cite{W}, adiabatic-passage techniques \cite{Linington08a,Linington08b}, or a HR-based technique \cite{Ivanov08}.
For Grover's search it is most convenient to use the following alternative \cite{Kyoseva06}:
 prepare the ion chain initially in state $\ket{\psi_0}\ket{1}$, and then apply, uniformly at all ions, a red-sideband pulse
 with a temporal area of $\pi/\sqrt{N}$ experienced by each ion (hence rms area of $\pi$).

2. {\em Grover iteration}
Each Grover logical step consists of two operators, an oracle call and a global HR \cite{Grover}.
Firstly, the oracle marks the searched state $\ket{\psi_m}$ by inverting its phase,
 which in our implementation is achieved by individual addressing of only the $m$th ion by a local $2\pi$ pulse;
 in fact, this is a HR operation $\mathbf{M}(\chi_m)$ with an interaction vector $\ket{\chi_m}$ identical to the basis vector $\ket{\psi_m}$ of Eq.~\eqref{k}.
The second HR $\mathbf{M}(\chi_W)$ with the HR vector $\ket{\chi_W} =\ket{\psi_W}$
 inverts the amplitude of all states $\ket{\psi_k}$ about the mean.
In our implementation this HR is produced by applying a single laser pulse, uniformly at all $N$ ions in the register, with a temporal area of $2\pi/\sqrt{N}$ (which amounts to rms-area of $2\pi$),
 as shown in Fig.~\ref{fig2}.
After the execution of the Grover operator $N_G$ times, the system is driven into the marked state,
 $\left[ \mathbf{M} (\chi_W) \mathbf{M}(\chi_m)\right] ^{N_G} \ket{\W} \approx \ket{\psi_m}$.

3. {\em Detection}
The marked ion is identified for it is the only one in state $\ket{1}$.

\begin{figure}[tb]
\includegraphics[angle=0,width=70mm]{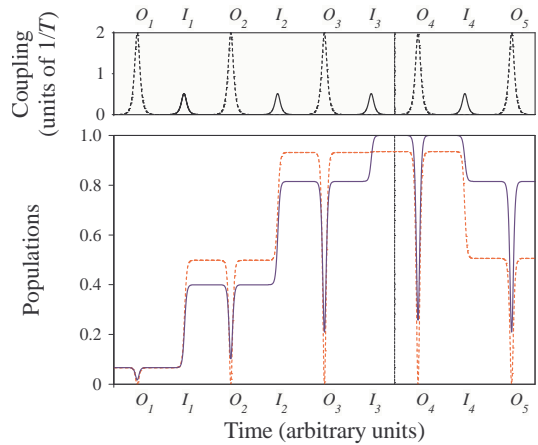}
\caption{(Color online) Numerically calculated population of the marked state vs time for $N=15$ ions for probabilistic (dashed line) and deterministic (solid line) Grover search.
The upper frame depicts the sequence of local (dashed, $O_k$, area $2\pi$) and global (solid, $I_k$, area $2\pi/\sqrt{N}$) pulses, both with sech shapes.
The individual couplings $g_{n}$ are given by the components of the QHR vectors $\ket{\chi_m}$ and $\ket{\chi_W}$,
each multiplied by $g=2/T$.
The detuning in the deterministic search is $\delta T\approx 0.589$, which produces the required phase $\varphi \approx 0.661\pi $.
Maximum probabilities occur after the completion of $I_3$, when the dashed curve reaches 0.92 and the solid line unity.}
\label{fig3}
\end{figure}

In Fig.~\ref{fig3} the probability to find the searched state for $N=15$ ions is plotted as a function of time.
In three interation steps, which amount to 3 local (oracle) pulses and 3 global (reflection) pulses, this probability increases to about 0.92, after which it decreases
 (as a part of oscillations between zero and unity in a long run).

{\bf \em Deterministic Grover search.}
The original Grover algorithm is probabilistic (except for $N=4$): it finds the marked state with a probability (fidelity) close to, but less than unity.
A slight modification of the original Grover algorithm, with the supplement of the (real) reflection operators by complex phase factors,
 makes the search fully deterministic, with a unit fidelity for any $N$, with $N_G$ or $N_G+1$ roundtrips \cite{Grover-complex}.

The deterministic search can be implemented with the same strategy as the original Grover search described above,
 by replacing the HRs $\mathbf{M}(\chi_m)$ and $\mathbf{M}(\chi_W)$ with the generalized HRs $\mathbf{M}(\chi_m;\varphi)$ and $\mathbf{M}(\chi_W;\varphi)$:
 $\left[ \mathbf{M}(\chi_W;\varphi) \mathbf{M}(\chi_m;\varphi) \right] ^{N_G}\ket{\W} = \ket{\psi_m}$.
The HR phase reads $\varphi = 2\sin^{-1}[\sqrt{N}\sin(\pi/(4N_G+6))]$ \cite{Grover-complex}; it can be produced by a suitable detuning $\delta$, as explained above.

In Fig. \ref{fig3} the population of the marked state is plotted for deterministic Grover search with $N=15$.
After just three iterations, the occupation probability for the marked state approaches unity, whereas it is only 0.92 for the standard Grover search.

{\bf \em Practical considerations.}
Now we discuss briefly various issues that may arise in a real experiment.
An obvious deviation from the idealized theory can be the imperfect initialization of the search register (the $W$-state), due to an inhomogeneous spatial profile of the driving laser(s),
 for example, if the laser beam for the global reflection is tilted from being collinear with the trap axis.
Then the initial state will not be an equal superposition, but the more general state $\ket{\Psi_a} = \sum_{n=1}^N a_n \ket{\psi_n}$,
 where the probability amplitudes  $a_n$ may deviate from $1/\sqrt{N}$, and they may also be complex, due to possible phase differences between the ions.
It has been shown that the register $\ket{\Psi_a}$ can still be used for Grover's search if the amplitudes $a_n$ do not deviate too much from $1/\sqrt{N}$ \cite{Grover98,Biham99}.
Biham \etal\ \cite{Biham99} have shown that the application of the same Grover operator, as for an equally weighted initial superposition, to the initial state $\ket{\Psi_a}$
 still produces the marked item with a reasonable probability.
However, it is readily shown that a HR with the same vector as the amplitude distribution in $\ket{\Psi_a}$,
\be\label{chi-unequal}
\ket{\chi} = [a_1,a_2,\dots,a_N]^T,
\ee
allows for a higher probability.
Using the approach of Hsieh and Li \cite{Hsieh} one can show that the optimum number of steps $N_G$ is replaced by $N_a=[\pi/(4|a_m|)]$ (for small $a_m$),
  where $a_m$ is the initial amplitude of the marked state $\ket{\psi_m}$.
Obviously, for $N_a$ to apply to \textit{any} state in the register $\ket{\Psi_a}$, the largest and smallest amplitudes in Eq.~\eqref{chi-unequal} should produce approximately the same $N_a$.

\begin{figure}[tb]
\includegraphics[angle=0,width=65mm]{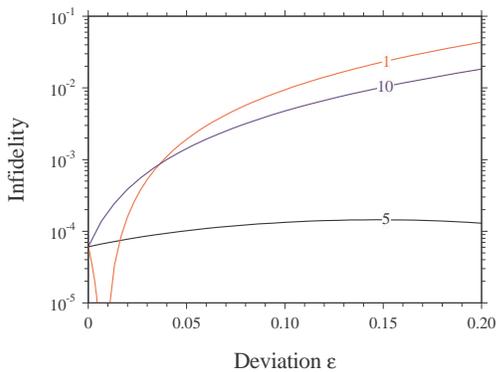}
\caption{(Color online) Infidelity of finding ions  1, 5 and 10  after 3 steps for $N=20$
 for a Gaussian spatial laser beam profile of the global pulse (rms area $2\pi$, detuning $\delta=0$),
 where the outside ions experience only a fraction $1-\varepsilon$ of the laser intensity in the middle of the ion chain.
}
\label{fig4}
\end{figure}

Figure \ref{fig4} shows the infidelity to detect the marked states  $\ket{\psi_1}$, $\ket{\psi_5}$ and $\ket{\psi_{10}}$ for $N=20$ ions after $N_G=3$ steps
 for a Gaussian laser beam profile of the global pulse,
 where the outside ions experience only a fraction $1-\varepsilon$ of the laser intensity in the middle of the ion string.
The proposed technique proves to be very robust with respect to the imperfectness of the laser profile.

In the proposed implementation the interaction vector \eqref{chi-unequal} can be produced very simply:
 by using the same laser beam as in the initialization step (which has produced the possible inhomogenuities in the register), but with twice as high Rabi frequency, cf. Fig.~\ref{fig2}.
This robustness to imperfections in the initial register, caused by unequal individual laser-ion couplings,
 also implies that the proposed implementation of Grover's search can utilize higher vibrational modes, e.g., the breathing mode,
 for which the laser-ion couplings depend on the position of the ion in the string \cite{James};
 using higher phonon modes greatly reduces deleterous heating effects.

{\bf \em Discussion and conclusions.}
In this Letter, we have proposed a very simple and concise implementation of Grover's search algorithm in a linear chain of $N$ trapped ions.
Unlike earlier proposals, which, in addition to the $\Osvet(\sqrt{N})$ queries, require many more physical nonquery steps,
 the proposed implementation requires only $\Osvet(\sqrt{N})$ such steps, one for each of the $\Osvet(\sqrt{N})$ queries;
 hence each Grover iteration is performed by one local HR (query) and one global HR (nonquery).
The speed-up in regard to physical steps derive from the natural emergence of the HR operator as the propagator in a red-sideband laser-driven linear string of trapped ions.

We point out that the proposed implementation, while using an entangled W-state as a quantum register,
 does not utilize the full power of a quantum register, which can contain $2^N$ states.
Work in this direction, which requires dealing with products of HRs, is in progress \cite{IIV-2}.

\acknowledgments

We acknowledge useful discussions with Ian Linington.
This work has been supported by the European Commission projects CAMEL, EMALI, and FASTQUAST, and the Bulgarian National Science Fund under grants VU-205/06 and VU-301/07.



\begin{thebibliography}{99}

\bibitem{Nielsen} M.A. Nielsen and I.L. Chuang, \emph{Quantum Computation and Quantum Information} (Cambridge University Press, Cambridge, 2000).

\bibitem{Grover} L.K. Grover, Phys. Rev. Lett. \textbf{79}, 325 (1997).

\bibitem{NMR-Grover}
 I.L. Chuang \etal, Phys. Rev. Lett. \textbf{80}, 3408 (1998);
 J.A. Jones \etal, Nature (London) \textbf{393}, 344 (1998);
 V.L. Ermakov and B.M. Fung, Phys. Rev. A \textbf{66}, 042310 (2002);
 J.E. Ollerenshaw \etal, Phys. Rev. Lett. \textbf{91} (2003);

\bibitem{NMR-8}
 L.M.K. Vandersypen and M. Steffen, Appl. Phys. Lett. \textbf{76}, 646 (2000).

\bibitem{Bhattacharya}
 N. Bhattacharya \etal, Phys. Rev. Lett. \textbf{88}, 137901 (2002).

\bibitem{Lloyd} S. Lloyd, Phys. Rev. A \textbf{61}, 010301 (1999).

\bibitem{Cirac-Zoller} J.I. Cirac and P. Zoller, Phys. Rev. Lett. \textbf{74}, 4091 (1995).

\bibitem{ion traps}
C. Monroe \etal, Phys. Rev. Lett. \textbf{75}, 4714 (1995);
Q.A. Turchette \etal, Phys. Rev. A \textbf{61}, 063418 (2000);
S. Gulde \etal, Nature (London) \textbf{421}, 48 (2003);
F. Schmidt-Kaler \etal, \ibid \textbf{422}, 408 (2003);
D. Leibfried \etal, \ibid \textbf{438}, 639 (2005).

\bibitem{W}
H. H\"{a}ffner \etal, Nature (London) \textbf{438}, 643 (2005).

\bibitem{Brickman}
 K.-A. Brickman \etal, Phys. Rev. A \textbf{72}, 050306 (2005);
 M. Feng, Phys. Rev. A \textbf{63}, 052308 (2001);
 C.D. Hill and H.-S. Goan, \ibid \textbf{69}, 056301 (2004).

\bibitem{Ivanov06}
 P.A. Ivanov \etal, Phys. Rev. A \textbf{74}, 022323 (2006).

\bibitem{Ivanov07}
 P.A. Ivanov \etal, Phys. Rev. A \textbf{75}, 012323 (2007).

\bibitem{Ivanov08} P.A. Ivanov and N.V. Vitanov, Phys. Rev. A \textbf{77}, 012335 (2008).

\bibitem{Kyoseva06} E.S. Kyoseva and N.V. Vitanov, Phys. Rev. A \textbf{73}, 023420 (2006).

\bibitem{Kyoseva08}
 E.S. Kyoseva \etal, J. Mod. Opt. \textbf{54}, 2237 (2008).

\bibitem{Grover02} L.K. Grover, Phys. Rev. A \textbf{66}, 052314 (2002).

\bibitem{James} D.F.V. James, Appl. Phys. B \textbf{66}, 181 (1998).

\bibitem{King}
 B.E. King \etal, Phys. Rev. Lett. \textbf{81}, 1525 (1998).

\bibitem{Linington08a} I.E. Linington and N.V. Vitanov, Phys. Rev. A \textbf{77}, 010302(R) (2008).

\bibitem{Linington08b} I.E. Linington and N.V. Vitanov, to be published.

\bibitem{Grover-complex} G.L. Long, Phys. Rev. A \textbf{64}, 022307 (2001).

\bibitem{Grover98}
 L.K. Grover, Phys. Rev. Lett. \textbf{80}, 4329 (1998).

\bibitem{Biham99} 	
E. Biham \etal, Phys. Rev. A \textbf{60}, 2742 (1999).

\bibitem{Hsieh}
J.-Y. Hsieh and C.-M. Li, \PRA \vol{65}, 052322 (2002).

\bibitem{IIV-2}
 S.S. Ivanov \etal, to be published.

\end{thebibliography}
\end{document}